# OPTICAL OBSERVATIONS OF THE BINARY MILLISECOND PULSARS J2145–0750 AND J0034–0534

J. F. Bell

*Mount Stromlo and Siding Spring Observatories, Institute of Advanced Studies, Australian National University, Private Bag, Weston Creek, ACT 2611, Australia*
*email: bell@mso.anu.edu.au*

S. R. Kulkarni

*Palomar Observatory, 105-24, California Institute of Technology, Pasadena, CA 91125*
*email: srk@astro.caltech.edu*

M. Bailes

*Australia Telescope National Facility, CSIRO, PO Box 76, Epping, NSW 2121, Australia*
*email: mbailes@atnf.csiro.au*

E. M. Leitch

*Palomar Observatory, 105-24, California Institute of Technology, Pasadena, CA 91125*
*email: eml@astro.caltech.edu*

A. G. Lyne

*Nuffield Radio Astronomy Laboratories, Jodrell Bank, University of Manchester, Macclesfield, Cheshire SK11 9DL, U. K.*
*email: agl@jb.man.ac.uk*

## ABSTRACT

We report on optical observations of the low-mass binary millisecond pulsar systems J0034–0534 and J2145–0750. A faint (I=23.5) object was found to be coincident with the timing position of PSR J2145–0750. While a galaxy or distant main-sequence star cannot be ruled out, its magnitude is consistent with an ancient white dwarf, as expected from evolutionary models. For PSR J0034–0534 no objects were detected to a limiting magnitude of R=25.0, suggesting that the white dwarf in this system is cold. Using white dwarf cooling models, the limit on the magnitude of the PSR J0034–0534 companion suggests that at birth the pulsar in this system may have rotated with a period as short as 0.6 ms. These observations provide further evidence that the magnetic fields of millisecond pulsars do not decay on time scales shorter than 1 Gyr.

*Subject headings:* Pulsars:individual PSR J0034-0534 PSR J2145-0750, Stars: binaries

## 1. Introduction

Millisecond pulsars are distinguished from normal pulsars by their rapid rotation rates and low magnetic field strengths. Considerable debate remains about their origin. In the standard model, the neutron star is spun up, accreting mass and angular momentum during the post-main-sequence evolution of a binary companion (Bhattacharya & van den Heuvel 1991). The standard model requires that the accretion process reduce the magnetic field strength from $10^{12}$ G, typical of the ordinary pulsars, to $\sim 10^{8}$ G, typical of millisecond pulsars. No physical models have been suggested that can successfully explain the magnetic field strengths of millisecond pulsars (Phinney & Kulkarni 1994). During this process the companion completes its post-main-sequence evolution and shrinks to form a He white dwarf. These white dwarfs are unusual with respect to the normal CO, 0.6 $M_\odot$ white dwarfs because they are composed of helium and have low masses ($\sim 0.2$ $M_\odot$).

Optical observations of these white dwarfs provide new insights into the formation and evolution of these systems (Kulkarni 1986, Callanan et al. 1989, Kulkarni, Djorgovski & Klemola 1991, Bell, Bailes & Bessell 1993, Bailyn 1993, Danziger, Baade & Della Valle 1993, Koester, Chanmugam & Reimers 1992, Nicastro et al. 1995, Lorimer et al. 1995a). For white dwarfs that have been detected in binary pulsar systems, white dwarf cooling theory offers independent estimates of the ages of such systems. From this work we now know that millisecond pulsars are long-lived objects and furthermore, that their magnetic fields do not decay on time scales shorter than 1 Gyr. However, these conclusions rest on the detection of only one companion, that of PSR J0437$-$4715 (Bell, Bailes & Bessell 1993) and several other sensitive but nonetheless, upper limits.

Searches for more companions are interesting for two further reasons. As mentioned above, the white dwarf companions are unusual compared to field white dwarfs and the physics of their cooling differs qualitatively from that of the ordinary white dwarfs. Also, assuming a quantitative cooling model, one can, in principle, infer the birth period of millisecond pulsars. This is a key parameter in evolutionary spin-up models and possibly provides an observational constraint on the equation of state of bulk nuclear matter (Kulkarni 1993).

The recent discovery of many millisecond pulsars in a survey of the southern sky with the Parkes radio telescope (Johnston et al. 1993, Bailes et al. 1994, Lorimer et al. 1995b) provided further opportunities to increase our understanding of these systems. The results and implications of optical searches for the companions of two binary millisecond pulsars J2145$-$0750 and J0034$-$0534 are the subject of this letter.

## 2. Optical Observations

Imaging data were obtained on the Palomar Hale 200-inch and the Australian National University 2.3-m telescopes in July 1993. The 2.3-m images through Cousins R and I filters were primarily used for astrometry and photometry. The COSMIC camera, at the prime focus of the Hale telescope, provided the deep images shown in Figure 1. through Gunn i and Cousins R filters. To find likely candidates for the companions, astrometry was performed using 10 or more stars from the Hubble Space Telescope Guide Star Catalog for each field. The positions of a number of candidate stars near the radio positions were also measured. Gunn i magnitudes were transformed to Cousins I magnitudes using i = 1.0084×I + 0.5427.

In the field of PSR J2145$-$0750, only one candidate star is within 3.5 arcsecs of the radio position, as shown in Figure 1. That star has I = 23.5 $\pm$ 0.1 and at 21 45 50.48 -07 50 18.02 it is 0.3 $\pm$ 1.0 arcsecs from the radio position. We consider it to be a plausible candidate for the companion. It is possible that this is a distant unrelated star but the probability of finding a main sequence star so distant at this Galactic latitude, $-72°$, is low. The probability of finding an unresolved faint red Galaxy is significant. However, the colour R$-$I $\simeq$ 0.6 is consistent with the object being a cool white dwarf and we proceed with the assumption that this is the pulsar companion.

Requiring a $3\sigma$ detection for PSR J0034$-$0534 we find no stars within several arcseconds of the radio position down to limiting magnitudes of I = 24.0 and R = 25.0. In the following section we discuss the implications of these observations for pulsar ages and initial rotation periods.

## 3. White Dwarf and Pulsar Ages

The rotational age of a pulsar, $\tau_p$ is defined to be:

$$\tau_p = \frac{P}{(n-1)\dot{P}_i}\left(1 - \left(\frac{P_0}{P}\right)^{n-1}\right) \qquad (1)$$

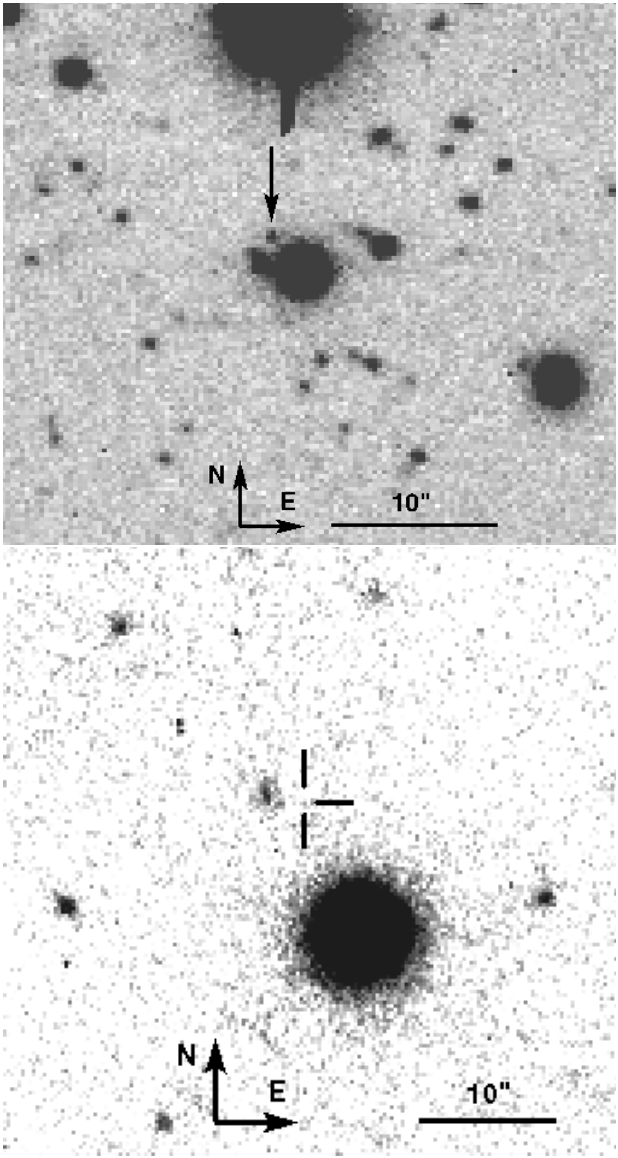

Fig. 1.— Top: Field around PSR J2145−0750 Gunn i. Bottom: Field around PSR J0034−0534 Kron-Cousins R. Pulsar positions are indicated with an arrow (J2145-0750) and cross hair (J0034-0534). Images in both were obtained with typical integration time of 800 s and later combined. The seeing was about 1 arcsecond (full width at half maximum) or better. The data were flat-fielded and analysed using the IRAF data reduction package.

where $n$ is the braking index of the pulsar, usually assumed to be 3, $P$ is the rotation period, $P_0$ is the rotation period at birth and $\dot{P}_i$ is the present value of the intrinsic period derivative of the rotation period. There are no measurements of braking indices for millisecond pulsars. For young pulsars $n \sim 2.5$ (Manchester & Taylor 1977), whereas $n = 3$ is predicted by the vacuum solution for a rotating magnetic dipole. If $P_0 << P$ and $n = 3$, then $\tau_c = P/2\dot{P}_i$ defines the characteristic age, which is regarded as an upper limit to the age of the pulsar. The observed $\dot{P}$ is corrupted by a kinematic effect $Pv^2/cd$ where $v$ is the transverse velocity and $d$ the distance (Shklovskii 1970, Camilo, Thorsett & Kulkarni 1994). At a distance of 0.5 kpc, the scintillation velocity of $31 \pm 25$ km s$^{-1}$ (Nicastro & Johnston 1995) for PSR J2145−0750 would contribute to the observed $\dot{P}$ at the 20% level. For PSR J0034−0534, $v$ has not been determined, but even a 50 km s$^{-1}$ velocity would contribute only 8% of the observed $\dot{P}$. Assuming a transverse velocity of 50 km s$^{-1}$ and the nominal distance estimates based on a model of interstellar electron distribution (Taylor & Cordes 1993) and the observed dispersion measures, the characteristic ages of PSRs J2145−0750 and J0034−0534 are 11.0 and 4.8 Gyr, respectively (see Table 1).

The most relevant white dwarf cooling model for the systems under consideration describes the cooling evolution of a 0.3 M$_\odot$ helium white dwarf in a binary system (Iben & Tutukov 1986). This model gives the bolometric luminosity $L$ of the white dwarf as a function of time. From Figure 2 of Iben and Tutukov, we obtain luminosities of $0.51 \times 10^{-4}$ L$_\odot$ and $1.20 \times 10^{-4}$ L$_\odot$ for PSR J2145−0750 and PSR J0034−0534, respectively. The cooling luminosity of a white dwarf for a given age scales directly with mass (Mestel 1952). Assuming that the companions only differ in mass from those in the cooling model, we scale the luminosities by 0.43/0.3 and 0.15/0.3, giving luminosities of $0.74 \times 10^{-4}$ L$_\odot$ and $0.60 \times 10^{-4}$ L$_\odot$ respectively. To predict magnitudes we assume the following: the spectra are well described by a blackbody, zero points given by Bessell (1979), the Stefan-Boltzmann law, $L = 4\pi r^2 \sigma T_e^4$ and the mass-radius relation for white dwarfs (Savonije 1983), $r = 0.013 M^{-1/3}$. For PSR J2145−0750, I $\leq$ 22.6, R $\leq$ 23.3 and T$_e \geq$ 4100 K, given a nominal distance of 0.5 kpc and an age of 10 Gyr. Similarly for PSR J0034−0534, I $\leq$ 24.5, R $\leq$ 25.4 and T$_e \geq$ 3250 K, given a nominal distance of 1.0 kpc and age of 4.8 Gyr.

Clearly the companion of PSR J2145−0750 (I = 23.5 ± 0.1) is fainter than the predicted magnitude of I ≤ 22.6. This indicates that either the companion has cooled faster than predicted by the white dwarf cooling models or the system is further away than the nominal distance of 0.5 kpc. Distance estimates based on dispersion measures are uncertain at the 30% level. Increasing the distance to 0.75 kpc would shift the predicted I magnitude to I ≤ 23.5, which agrees with the observed value. The companion was also detected in a deep image, R = 24.1 ± 0.4. This is consistent with the R–I colour predicted above. So far we have assumed that the mass of the companion is around the minimum possible, that is 0.43 $M_\odot$. The equation of state for these white dwarfs is somewehat uncertain, making it difficult to estimate the uncertainties in the cooling age. Changing the mass of the companion to $m_c = 0.6$ $M_\odot$ and using the cooling curve for a CO white dwarf does not affect the predicted magnitudes, but increases the inferred effective temperature to ≥ 4300 K. With $M_v \simeq 15.8$, this white dwarf is amongst the coldest and oldest known (Monet et al. 1992, Winget et al. 1987, Kapranidis 1985).

## 4. Initial Rotation Period of PSR J0034−0534

The companion of PSR J0034−0534, at the predicted effective temperature of ≥ 3250 K, has R–I ≃ 0.9. Since the companion can only be younger and therefore hotter than 3250 K, the R–I must be ≤ 0.9. Hence, the limit of R = 25.0 is tighter than I = 24.0, when compared with the predicted magnitudes of R ≤ 25.4 and I ≤ 24.5. Comparing the limiting magnitude of 25 in R and the predicted R ≤ 25.4, the cooling age is $\tau_c$ >4.0 Gyr.

PSR J0034−0534 is one of the most rapidly rotating pulsars known ($P = 1.877$ ms). Using the 4.0 Gyr cooling age in Equation 1, with $n = 3$, we derive $P_0 = 0.6$ ms. Several of the parameters used to derive this value are uncertain. If $m_c > 0.15$ $M_\odot$ or $d < 1.0$ kpc or $\tau_c > 4.0$ Gyr or $n > 3$ or $v < 50$ km s$^{-1}$, an even smaller value of $P_0$ is implied. The converse is of course true if the above inequalities are reversed.

Taking an extreme case, with $d = 1.25$ kpc, minimum companion mass, and a transverse velocity of 100 km s$^{-1}$, a cooling age at the more rapid end of the possible range, and a low braking index of 2.5, we find $P_0 < 1.6$ ms, near the current spin periods of PSR B1957+20 and PSR B1937+21. It is unlikely that *all* of these values are so far from those more normally assumed. We therefore conclude that the initial spin period of PSR J0034−0534 could easily have been much less than 1.6 ms.

This is a fascinating possibility, given that the theoretical minimum rotation periods for neutron stars are $0.3 - 1.6$ ms depending on the equation state. The observational constraints on a class of sub-millisecond pulsars is weak since all pulsar searches to date have not had significant sensitivity to sub-millisecond pulsars (Kulkarni 1993, Lyne 1995). The accuracy to which $P_0$ can be determined will improve with measurement of the proper motion, deeper optical searches and continued precision timing.

We thank M. Bessell and the referee D. Winget for helpful suggestions. M. Bailes is a Queen Elizabeth II Fellow and J. Bell received support from an Australian Postgraduate Research Award and the Australia Telescope National Facility student program. S. Kulkarni's work is supported by grants from NASA, NSF and the Packard Foundation.

---

| Pulsar Name | J0034−0534 | J2145−0750 |
| --- | --- | --- |
| Right Ascension (J2000) | $00^{\rm h}34^{\rm m}21\overset{\rm s}{.}826(2)$ | $21^{\rm h}45^{\rm m}50\overset{\rm s}{.}468(2)$ |
| Declination (J2000) | $-05°34'36\overset{''}{.}56(3)$ | $-07°50'18\overset{''}{.}29(8)$ |
| Galactic Latitude (deg) | −68 | −42 |
| Galactic Longitude (deg) | 111 | 47 |
| Period (ms) | 1.877181854020(2) | 16.05242365500(8) |
| Period derivative | $0.7(1) \times 10^{-20}$ | $2.9(2) \times 10^{-20}$ |
| Period epoch (MJD) | 48765.986000 | 48978.657300 |
| Dispersion Measure ($cm^{-3}$ pc) | 13.763(2) | 9.043(2) |
| Orbital period (days) | 1.58928180(3) | 6.83890247(4) |
| Semi-major axis (lt-s) | 1.437768(5) | 10.16409(2) |
| Eccentricity | 0.0000(1) | 0.000018(4) |
| Epoch of periastron (MJD) | 48765.6(1) | 48932.3(2) |
| Longitude of periastron (deg) | 0(assumed) | 206(12) |
| Characteristic age (Gyr) | 4.8(7) | 11.0(7) |
| Distance (kpc) | 1.0(3) | 0.5(2) |
| Mass function ($M_\odot$) | 0.0012634(3) | 0.0241054(2) |
| Min. Companion Mass ($M_\odot$) | 0.15 | 0.43 |
| R magnitude | >25.0 | 24.1(4) |
| I magnitude | >24.0 | 23.5(1) |
| Integration time R (s) | 4000 | 2400 |
| Integration time I (s) | 2400 | 4600 |

Table 1: Parameters for PSR J2145−0750 and PSR J0034-0534. The numbers in brackets signify the $2\sigma$ uncertainty in the last quoted digit.